\documentclass[prb,twocolumn]{revtex4}
\usepackage{graphicx}

\begin{document}

\title{Phase diagram of imbalanced fermions in optical lattices}
\author{Xiaoling Cui and Yupeng Wang}
\affiliation{Beijing National Laboratory for Condensed Matter
Physics and Institute of Physics, Chinese Academy of Sciences,
Beijing 100190, China
\\}
\date{{\small \today}}
\begin{abstract}

The zero-temperature phase diagrams of imbalanced fermions in 3D
optical lattices are investigated to evaluate the validity of the
Fermi-Hubbard model. It is found that depending on the filling
factor, $s$-wave scattering strength and lattice potential, the
system may fall into the normal($N$) phase, magnetized
superfluid(SF$_M$) or phase separation of $N$ and BCS state. By
tuning these parameters, the superfluidity could be favorable by
enhanced effective couplings or suppressed by the increased band
gap. The phase profiles in the presence of a harmonic trap are also
investigated under LDA, which show some exotic shell structures
compared to those without the optical lattice.

\end{abstract}

\maketitle

In the past few years, great experimental progress has been achieved
in studying ultracold Fermi gases with polarization
\cite{zwierlein06,partridge06,zwierlein062,schunck07,shin07}. With
two unequal mixtures of cold $^6 $Li atoms in a harmonic
trap\cite{zwierlein06,partridge06}, a clear evidence of phase
separation with an unpolarized superfluid (BCS) core and a normal
($N$) shell around that has been observed in experiment.
Theoretically
\cite{sarma63,fflo64,bedaque03,sheehy06,gu06,pao06,hu06,parish07},
many other ground state candidates have been proposed in such
systems, including magnetized superfluid (SF$_M$), phase separation
(PS), and Fulde-Ferrell-Larkin-Ovchinnikov (FFLO) state with finite
momentum pairing by tuning  the interaction parameter $1/k_Fa_s$,
the polarization $P=\delta n/n$ or Zeeman field $h$. Since the
optical lattice height $V_0$ is  also tunable, it is very
 interesting to study its effect on the new phase diagram. For equal mixtures,
 a second-order quantum
phase transition between superfluid (SF) and insulating (IN) phases
has been addressed both experimentally at a critical lattice height
$V_c$ at resonance \cite{chin06} and
theoretically\cite{zhai07,moon07} based on the second-order
perturbation theory. Besides, there are also works on imbalanced
fermions in optical lattices focusing on IN\cite{iskin07},
FFLO\cite{koponen07} and SF$_M$\cite{dao08} phases, based on an
effective Fermi-Hubbard model.

In this work, starting from the exact lattice spectrum, we study the
ground state phase diagram of imbalanced two species Fermi gases
trapped in 3D optical lattices, in terms of the total filling factor
$n$, polarization $P$, $s$-wave scattering length $a_s$ and lattice
potential $V_0$. Limited by the numerical attainment, the FFLO-type
pairing is not considered. The total pairing reciprocal lattice
momentums involved in our simulation are up to the six smallest
non-zero ones, which turn out to be more and more important as $V_0$
increases. Sufficient multiple bands have been taken into account to
ensure the accuracy especially in the strong coupling regime. We
demonstrate that there are two contradictory effects of $V_0$ on the
SF phase, depending on the average filling factor $n$. One is the
enhanced density of states (DOS) inside each band which effectively
increases the coupling strength and thus is favorable to SF; the
other is the broadened band gap or discontinuity of DOS which is
against SF. One key point is that besides tuning $a_s$ through the
Feshbach resonance(FR), $V_0$ can also be tuned and drive the system
from weak to strong coupling regime, provided that the filling
factor is properly fixed. An obvious evidence is the emergence of
SF$_M$ phase for deep optical lattices at particular filling
regimes, even in far BCS side of FR. We also propose that the
critical polarization versus total filling factor diagram obtained
can be used to evaluate the validity of the usual Fermi-Hubbard
model. The phase profile in the presence of an external harmonic
trap, which is more relevant to the practical experiment will be
studied with local density approximation (LDA) finally. Some exotic
structures appear, reflecting the uniqueness of the optical
lattices.

In a recent experiment\cite{chin06}, two hyperfine states of
ultracold $^6$ Li atoms, $|F=1/2,m_F=1/2\rangle$
($|\uparrow\rangle$) and $|F=1/2,m_F=-1/2\rangle$
($|\downarrow\rangle$), had been successfully loaded to an optical
lattice. The low-energy interactions are characterized by a single
$s$-wave scattering length $a_s$, which can be tuned by FR. Such a
system can be well described by  the one-channel Hamiltonian
\begin{eqnarray}
H&=&\int d\mathbf{r}\sum_{\sigma=\uparrow,\downarrow}
\psi^{\dag}_{\sigma}(\mathbf{r})\hat{H}_0(\mathbf{r})\psi_{\sigma}(\mathbf{r})+\nonumber\\
&&g \int d\mathbf{r}
\psi^{\dag}_{\uparrow}(\mathbf{r})\psi^{\dag}_{\downarrow}(\mathbf{r})\psi_{\downarrow}(\mathbf{r})\psi_{\uparrow}(\mathbf{r}),\label{hamil}
\end{eqnarray}
where $\hat{H}_0=\sum_{i=x,y,z}-\hbar^2\partial_i^2/2M+V_0\sin^2(\pi
x_i/a)$; $a=\lambda/2$ is the period of the lattice generated in
each direction by two oppositely propagating lasers with wavelength
$\lambda$; $V_0$ is the lattice height which is usually measured by
the recoil energy $E_R=\frac{\hbar^2\pi^2}{2Ma^2}$; $g$ is the
renormalized contact interaction constant between two species by
eliminating the unphysical divergence due to the high-momentum
contribution for fermi gases,
$\frac{1}{g}=\frac{m}{4\pi\hbar^2a_s}-\frac{1}{V}\sum_q
\frac{1}{2\epsilon_q}$.

In the framework of mean-field approach, we expand first each field
operator in terms of eigenwave functions of $\hat{H}_0$,
$\psi_{\sigma}(\mathbf{r})=\sum_{\mathbf{nk}}\phi_{\mathbf{nk}}(\mathbf{r})
\psi_{\mathbf{nk}\sigma}$. The Bloch wave functions
$\phi_{\mathbf{nk}}(\mathbf{r})=\frac{1}{\sqrt{V}}\sum_{\mathbf{G}}
a_{\mathbf{nk}}(\mathbf{G})e^{i(\mathbf{k}+\mathbf{G})\cdot\mathbf{r}}$
and energies $\epsilon_{\mathbf{nk}}$ are obtained from the
Schr{\"o}dinger equation
\begin{eqnarray}
\sum_{\mathbf{G'}}\Big\{[\frac{\hbar^2}{2M}(\mathbf{k+G})^2&+&\frac{3V_0}{2}]\delta_{\mathbf{GG'}}-\frac{V_0}{4}\sum_i\delta_{\mathbf{G}\pm\frac{2\pi}{a}\mathbf{e}_i,\mathbf{G'}}\Big\}\nonumber\\
&&a_{\mathbf{nk}}(\mathbf{G'})=\epsilon_{\mathbf{nk}}a_{\mathbf{nk}}(\mathbf{G}),
\end{eqnarray}
where $\mathbf{n}=\{n_x,n_y,n_z\}=s,p,...$ indicate the band
indices; $\mathbf{k}$ lie in the first Brillouin zone (BZ) and
$\mathbf{G}=2\pi/a(l_x,l_y,l_z)$ is the reciprocal lattice vector.
The solutions satisfy
$\sum_{\mathbf{G}}a_{\mathbf{nk}}^*(\mathbf{G})a_{\mathbf{n'k}}(\mathbf{G})=\delta_{\mathbf{nn'}}$
and $a_{\mathbf{n,-k}}(\mathbf{-G})=a_{\mathbf{nk}}^*(\mathbf{G})$.
The standard mean-field treatment gives
\begin{eqnarray}
H-&&\sum_{\sigma}\mu_{\sigma}N_{\sigma}=\sum_{\mathbf{nk}\sigma}(\epsilon_{\mathbf{nk}}-\mu_{\sigma})
\psi^{\dag}_{\mathbf{nk}\sigma}\psi_{\mathbf{nk}\sigma}-\nonumber\\
&&\sum_{\mathbf{mnk}}(\Delta_{\mathbf{mnk}}^*\psi_{\mathbf{m -k}
\downarrow}\psi_{\mathbf{nk}\uparrow}+h.c.)-\frac{V}{g}\sum_{\mathbf{Q}}
|\Delta_\mathbf{Q}|^2,\label{hamil2}
\end{eqnarray}
with
\begin{eqnarray}
\Delta_{\mathbf{Q}}&=&-\frac{g}{V}
\sum_{\mathbf{mnk}}M^{\mathbf{Q}}_{\mathbf{mnk}}\langle
\psi_{\mathbf{m -k}\downarrow}\psi_{\mathbf{nk}\uparrow}\rangle ,\nonumber\\
\Delta_{\mathbf{mnk}}&=& \sum_{\mathbf{Q}}\Delta_\mathbf{Q}
M^{\mathbf{Q} \ *}_{\mathbf{mnk}},\label{delt}
\end{eqnarray}
and $M^{\mathbf{Q}}_{\mathbf{mnk}}=\sum_{\mathbf{G}}a_{\mathbf{m
-k}}(\mathbf{-G})a_{\mathbf{nk}}(\mathbf{G+Q})$. Since
$M^{\mathbf{Q}=0}_{\mathbf{mnk}}=\delta_{\mathbf{mn}}$ and if $m\neq
n$ $M^{\mathbf{Q}\neq0}_{\mathbf{mnk}}$ are quite small, in the
following text we only consider pairing within each single band,
which means
$\Delta_{\mathbf{mnk}}\approx\Delta_{\mathbf{nk}}\delta_{\mathbf{mn}}$.
In such a case, the Hamiltonian can be easily diagonalized, and the
thermodynamic potential is calculated at $T=0$ as
\begin{eqnarray}
\frac{\Omega}{V}&=&\frac{1}{V}\sum_{\mathbf{nk}}\{\Theta(-E_{\mathbf{nk}+})E_{\mathbf{nk}+}+\Theta(-E_{\mathbf{nk}-})E_{\mathbf{nk}-}+\nonumber\\
&&\epsilon_{\mathbf{nk}}-\mu-\sqrt{(\epsilon_{\mathbf{nk}}-\mu)^2+\Delta_{\mathbf{nk}}^2}\}-\sum_{\mathbf{Q}}
\frac{|\Delta_{\mathbf{Q}}|^2}{g},\label{omega}
\end{eqnarray}
with
$E_{\mathbf{nk}\pm}=\sqrt{(\epsilon_{\mathbf{nk}}-\mu)^2+\Delta_{\mathbf{nk}}^2}\mp
h$ where $\mu=(\mu_{\uparrow}+\mu_{\downarrow})/2$ and
$h=(\mu_{\uparrow}-\mu_{\downarrow})/2$. From
$\partial\Omega/\partial\Delta_{\mathbf{Q}}^*=0$ and
$N_{\sigma}=-\partial\Omega/\partial\mu_{\sigma}$, we get the gap
and density equations as
\begin{eqnarray}
-\frac{\Delta_{\mathbf{Q}}}{g}&=&\frac{1}{V}\sum_{E_{\mathbf{nk}\pm}>0}\frac{M_{\mathbf{nk}}^{\mathbf{Q}}
\Delta_{\mathbf{nk}}}{2\sqrt{(\epsilon_{\mathbf{nk}}-\mu)^2+\Delta_{\mathbf{nk}}^2}},\\\label{gap-eq}
n&=&\frac{1}{N_L}(\sum_{\mathbf{nk}}1-\sum_{E_{\mathbf{nk}\pm}>0}
\frac{\epsilon_{\mathbf{nk}}-\mu}{\sqrt{(\epsilon_{\mathbf{nk}}-\mu)^2+\Delta_{\mathbf{nk}}^2}}),\nonumber\\
\delta
n&=&\frac{1}{N_L}(\sum_{E_{\mathbf{nk}+}<0}1-\sum_{E_{\mathbf{nk}-}<0}1).\label{den-eq}
\end{eqnarray}
Here $N_L$ is the total number of lattice sites,
$n=(N_\uparrow+N_\downarrow)/N_L$ and $\delta
n=(N_\uparrow-N_\downarrow)/N_L$ are the total filling factor and
the difference, respectively. Hereafter we scale all the energies in
unit of $E_R$ and the momenta of $2\pi/a$.

To make the numerical simulations attainable but still retain the
essence of the problem, besides $\mathbf{Q}=0$ we consider other six
non-zero $\mathbf{Q}$: $(\pm1,0,0),(0,\pm1,0),(0,0,\pm1)$. Due to
$M_{\mathbf{nk}}^{\mathbf{Q}}=M_{\mathbf{n \pm k}}^{\mathbf{\pm Q}}$
and the isotropy of 3D cubic lattices, all six non-zero $\mathbf{Q}$
share the same pairing amplitude $\Delta_1$. Therefore we get two
coupled gap equations in terms of $\Delta_0$ and $\Delta_1$. For a
realistic numerical simulation, we apply a cutoff momentum
$|q_{\Lambda}|=\frac{3}{2}(1,1,1)$ in the renormalization and
correspondingly consider the lowest three bands($s,p,d$) in each
direction of lattice spectrum. This truncation allows totally $n=54$
atoms per site at most, which is well above the experimental
interest as well as ours in this paper.

\begin{figure}[ht]
\includegraphics[height=6cm]{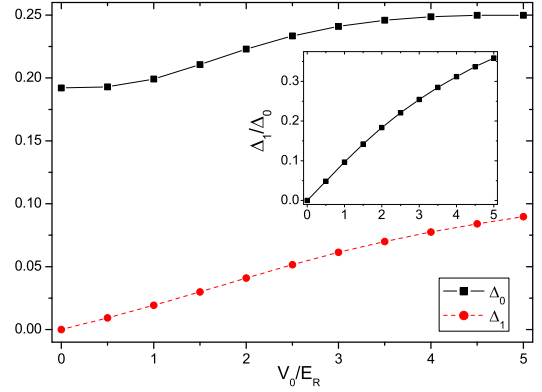}
\caption{(color online) $\Delta_0$, $\Delta_1$ and
$\Delta_1/\Delta_0$(see inset) vs lattice potentials $V_0$ for equal
mixtures. $a/a_s=-3$. The averaged filling is fixed to be $n=1$.}
\label{n1Delt}
\end{figure}

Before turning to the phase diagram of imbalanced system, first we
analyze the necessity of involving non-zero $\mathbf{Q}$ in gap
equations for equal mixtures. Fig. \ref{n1Delt} shows
$\Delta_0$,$\Delta_1$ and their ratio as a function of lattice
potential $V_0$ at half filling $n=1$. It is shown that the
$\mathbf{Q}\neq0$ pairing becomes more and more important as $V_0$
increases. This effect can be understood as follows. Taking a very
shallow 1D lattice for example, non-zero $M_{nk}^Q$ with
$|Q|=2,4,6...$ and $1,3,5...$ only exist around kinetic
energy-degenerate points $k=0$ and $k=\pm1/2$ respectively, which
contribute little to gap-equations and therefore produce a
negligible $\Delta_1$. In the limit of $V_0=0$, these non-zero
$M_{nk}^Q$ exactly cancel with each other in gap equations and
finally only $Q=0$ pairing survives. However as $V_0$ increases, the
eigen-vector ${a_{nk}(G)}$ evolves such that the area of non-zero
$M_{nk}^Q$ expands from three discrete points in first BZ to
considerable regions around them, leading to an increasing
$\Delta_Q$ with $Q\neq0$. Since our interest is still within
$s$-band, the $|Q|=1$ pairings take a leading role among all
non-zero ones, which is verified both numerically and analytically
from a perturbation theory for shallow lattices. This is why we just
take into account six smallest non-zero $\mathbf{Q}$ in $3D$ case
for not-so-deep lattices. The consideration of non-zero
$\mathbf{Q}$-pairing would produce a much stronger superfluidity
especially for deep lattices, which can also be seen from the
comparison of the previous two works\cite{compare}.

\begin{figure}[ht]
\includegraphics[width=\columnwidth]{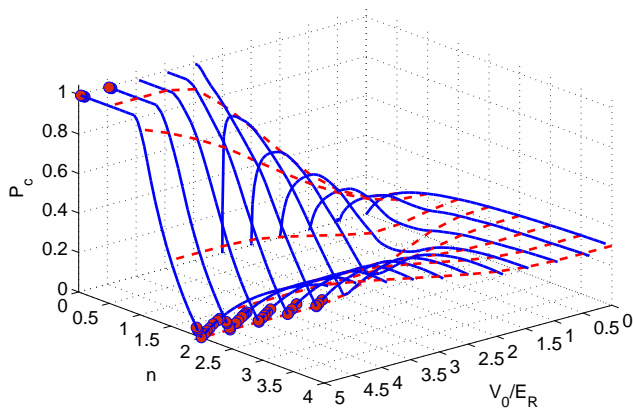}
\caption{(color online) Zero temperature phase diagram as a function
of polarization $P_c=\delta n/n$, total filling factor $n$ and
lattice height $V_0$. $a/a_s=-3$. The red dashed(blue solid) lines
show $P_c$ evolves with $V_0$($n$) for fixed $n$($V_0$). All the
lines above denote the $PS-N$ boundaries, expect for red solid
circles separating SF$_M$ and $N$ phase instead. } \label{3d}
\end{figure}

The ground state phase diagram in Fig. \ref{3d} is determined as
follows. We first compare
$\Omega_{BCS}(\mu,h=E_{min},\Delta=\{\Delta_{0},\Delta_{1}\})$ with
with $\Omega_{N}(\mu,h=E_{min},\Delta=\{0\})$, with $\Delta_{0/1}$
obtained for unpolarized BCS state and $E_{min}={\rm
Min}_{nk}(\sqrt{(\epsilon_{nk}-\mu)^2+\Delta_{nk}^2})$ its lowest
excitation energy. If $\Omega_{BCS}>\Omega_{N}$ then the first-order
phase transition point $h_c(<E_{min})$ is given by
\begin{eqnarray}
\Omega_{BCS}(\mu,h_c,\Delta=\{\Delta_{0},\Delta_{1}\})&=&\Omega_{N}(\mu,h_c,\Delta=\{0\}),\\
n_N(\mu,h_c)&=&n.
\end{eqnarray}
$P_c=\delta n_N(\mu,h_c)/n$ represents a critical point when PS is
entirely composed by $N$ phase. Note that in this case the polarized
SF or Sarma phase\cite{sarma63} is unstable due to the negative
superfluid density\cite{pao06}. If $\Omega_{BCS}<\Omega_{N}$ then
there should be a stable SF$_M$ interpolating between BCS and $N$
phase. In free space at the SF$_M$-N second-order transition point,
$N$ denotes a fully polarized normal state with
$P_c=1$\cite{parish07}. Correspondingly in optical lattices, we
obtain $P_c=1$ at $n\ll 1$ and $|n-2|/n$ at $n\sim2$, as shown by
red solid circles in Fig. \ref{3d}.

\begin{figure}[ht]
\includegraphics[height=6cm]{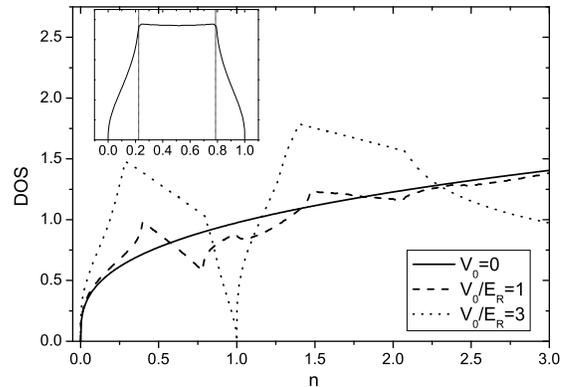}
\caption{Density of state(DOS) at the Fermi surface versus filling
factor $n$ for single-spin atoms in 3D free space and in lattices
with $V_0/E_R=1$(no band gap) and $V_0/E_R=3$(with band gap). Inset
is DOS for non-interacting Hubbard model. The dotted lines therein
denote two peaks of DOS at $(\mu=t,n_p=0.213)$ and
$(3t-\mu,1-n_p)$.} \label{dos}
\end{figure}

We analyze that $P_c-V_0$ curves reveal two contradictory effects of
increasing $V_0$ to SF depending on the filling factor $n$. As shown
in Fig. \ref{dos}, for $n\leq1$ increasing $V_0$ will flatten each
band and enhance DOS(almost inversely proportional to the band
width); while at $n\sim2$, increasing $V_0$ produces an entirely
opposite effect due to the enlarged band gap. According to the
standard BCS theory, the DOS at the Fermi surface dramatically
affects the strength of SF, as is also reflected by such
contradictory effects. When $V_0=3E_R$, $P_c$ increases to unity at
small $n$ but reduces to zero at $n=2$, denoting the IN phase with
$n_{\uparrow}=n_{\downarrow}=1$. For $n\in(1,2)$, $P_c$ initially
drops down and then goes up, indicating the competition between the
above these two effects. Here the lattice enhancement of $P_c$ at
$n\leq1$ is similar to the enhancement of $T_c$ for equal mixtures
in weak coupling limit\cite{micnas90,hofstetter02}.

Next we turn to $P_c-n$ diagram for fixed $V_0$. As is well known in
free space, a first-order BCS to $N$ phase transition takes place in
weak coupling limit at $h_c=\frac{\Delta_0}{\sqrt{2}}$ and
$P_c=\frac{3h_c}{2E_F}$, with the gap amplitude
$\Delta_0=\frac{8}{e^2} E_Fexp(-\frac{\pi}{2k_F|a_s|})$ and the
interaction parameter
$\eta=\frac{1}{k_Fa_s}=\frac{a}{a_s}(3\pi^2n)^{-\frac{1}{3}}$. $P_c$
will increase with $n$ all along from weak coupling
limit($\eta\rightarrow -\infty$) to the unitary
limit($\eta\rightarrow 0$). Within an optical lattice, however, the
$P_c-n$ curve would be dramatically modified. In weak coupling limit
with small $V_0$, the curve basically follows as that of DOS in Fig.
\ref{dos}, with a dip at $n_d\sim2$ and correspondingly a peak at
$n_p$. As $V_0$ increases, $n_p$ gradually moves to the left and
finally vanishes at $n_p=0$, and finally SF$_M$ state emerges at
$n\ll1$ or $n\sim2$. Different from the SF$_M$ studied by DMFT
method under tight-binding model\cite{dao08}, the phase shown here
is purely due to the enhanced effective coupling by lattices. In
this limit, two fermions are likely to form a molecule, and the BCS
equation directly reduces to a Schrodinger equation for a single
bound pair\cite{nozieres85,micnas90}. It is expected that as $V_0$
increases, the SF$_M$ phase would extend to a larger or even the
whole density region. Actually, the physics at $n\ll 1$ and $n\sim2$
can be related to each other via particle-hole symmetry. The
symmetry is essentially obvious within the background of
Fermi-Hubbard model,
$\epsilon_{\mathbf{k}}=0.5t\sum_i(1-\cos(k_ia))$. Since $(n,\mu)$
and $(2-n,3t-\mu)$ share the same $\{\Delta,h,\delta n, \Omega\}$
and thus the same critical $\delta n_c$, the critical polarization
$P_c(n)$ for $n\leq8$ follows as
\begin{equation}
P_c=\left\{\begin{array}{cc}
             1, & n\leq1 \\
             \frac{|n-2|}{n}, & 1<n\leq5 \\
             \frac{8-n}{n}. & 5<n\leq8
           \end{array}
 \right.
\end{equation}

\begin{figure}[ht]
\includegraphics[height=6cm]{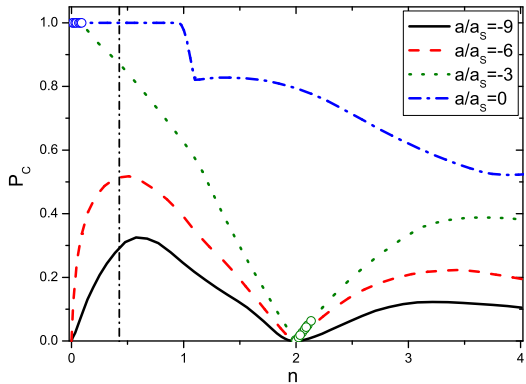}
\caption{(color online) $P_c$ versus $n$ diagram at different
couplings. $V_0=3E_R$. All lines denote $PS-N$ boundaries, except
that the green and blue circles show $SF_M-N$ boundaries. The black
dashed-dot line marks the upper limit of peak position based on
Fermi-Hubbard model(see text). } \label{asv3}
\end{figure}

We also compute the phase diagram at other $s$-wave couplings with
fixed $V_0$ as shown in Fig. \ref{asv3}. Different from Fig.
\ref{3d}, it shows that the increasing $a/a_s$ always enhance SF and
improve $P_c$ regardless of filling factors. At sufficiently strong
coupling close to unitary, the particle-hole symmetry in each band
breaks down since it is energetically favorable for particles in
s-band to overcome the band-gap and form cooper pairs even at $n=2$.
In this case the multi-band effect should be taken into account.
This is why SF$_M$ only turns up at $n\ll1$ but not at $n\sim2$ in
unitary limit, shown as blue circles in Fig. \ref{asv3}.

We emphasize that the $P_c-n$ diagram in weak coupling limit can be
used to evaluate the validity of tight-binding approximation(TBA)
usually applied to the lattices. For Hubbard model, the DOS shows
two peaks symmetric around half filling(see inset of Fig.
\ref{dos}), due to the van Hove singularity at $(\mu=t,n=0.213)$ and
$(\mu=2t,n=0.787)$. We also verified numerically that the peak
position of $P_c$ at different couplings is never greater than
$0.426$ for arbitrary interactions $|U|/t$, twice as that for the
first peak in DOS. This universal property could be used to justify
the validity of TBA to realistic lattices. Apparently from Fig.
\ref{asv3} we see that the TBA is not applicable to $V_0=3E_R$,
since at really weak coupling($a/a_s=-9$) the peak position
$n_p\simeq0.6>0.426$. The disagreement here indicates the deviation
of the two lattice spectrum, and thus the necessity of adopting
exact lattice spectrum for not-so-deep lattices.

Finally, it is also useful to consider the phase profile relevant to
realistic experiments with an external harmonic potential
$V(\mathbf{r})$. Under LDA, the system is assumed to be locally
homogeneous with an averaged chemical potential
$\mu(\mathbf{r})=(\mu_{0\uparrow}+\mu_{0\downarrow})/2-V(\mathbf{r})$
and position-independent difference
$h=(\mu_{0\uparrow}-\mu_{0\downarrow})/2$. The phase at position
$\mathbf{r}$ is determined by the local $(\mu(\mathbf{r}),h)$, which
is also self-consistently related to the total particle numbers,
$s$-wave interaction and the lattice potential. Here we give several
typical phase profiles with the filling factor in trap center larger
than $2$. For relatively shallow lattices and very weak $s$-wave
interactions, a typical one from the trap center to the edge is:
BCS-PN-IN-PN-BCS-PN-FN (PN/FN: partially/fully polarized Normal).
Starting from this profile, if $V_0$ increases, PN is very likely to
be replaced by SF$_M$ at positions where $n(\mathbf{r})\sim2$ or
$n(\mathbf{r})\ll1$, while IN still survive in a certain region; But
if $s$-wave interaction increases, then IN shrinks gradually, two
BCS regimes merge together and PN gives rise to SF$_M$, with only
three phases left finally: BCS-SF$_M$-FN. In the latter case, much
higher bands with continuous spectrum would be occupied, which makes
the situation very similar to free space and the lattice effect is
not obvious.

We thank Fei Zhou and An-Chun Ji for beneficial discussions. X.L.C.
would like to thank the hospitality of UBC during her stay in
Canada, where this work was initially started. This work is in part
supported by NSFC under Grant No. $10574150$ and
$973$-Project(China) under Grant No. $2006CB921300$.


\begin{thebibliography}{}
\bibitem{compare}Comparing two works on $SF-IN$
transition for equal mixtures at resonance, the critical lattice
potential $V_c$ calculated in one\cite{moon07} is almost one order
of magnitude larger than in the other\cite{zhai07}, due to the
consideration of couplings between different reciprocal lattice
vectors.
\end{thebibliography}
\end{document}